# Standardized Analysis Ready (STAR) data cube for high-resolution Flood mapping using Sentinel-1 data

Surajit Ghosh[1], Arpan Dawn[2], Sneha Kour[2] and Susmita Ghosh[3]

[1]International Water Management Institute, Sri Lanka
[2]The University of Burdwan, India
[3]Jadavpur University, India

Contact: s.ghosh@cgiar.org

*Summary*
Floods are one of the most common disasters globally. Flood affects humans in many ways [1, 2, 3]. Therefore, rapid assessment is needed to assess the effect of floods and to take early action to support the vulnerable community in time. Sentinel-1 is one such Earth Observation (EO) mission widely used for mapping the flooding conditions at a 10m scale [4]. However, various preprocessing steps are involved before analyses of the Sentinel-1 data. Researchers sometimes avoid a few necessary corrections since it is time-consuming and complex. Standardization of the Sentinel-1 data is the need of the hour, specifically for supporting researchers to use the Standardized Analysis-Ready (STAR) data cube without experiencing the complexity of the Sentinel-1 data processing. In the present study, we proposed a workflow to use STAR in Google Earth Engine (GEE) environment. The Nigeria Flood of 2022 has been used as a case study for assessing the model performance.

## I. INTRODUCTION

Through its Copernicus program, the European Space Agency (ESA) launched the Sentinel-1A and 1B satellites in March 2014 and 2016, respectively. The orbital period of the two Sentinel-1 satellites is once every 12 days, allowing for a combined potential repeat frequency of 6 days over the equator and a revisit frequency of 3 days when both ascending and descending orbits are considered. Four imaging modes are operationally used to collect data from the Sentinel-1 satellites: interferometric wide-swath (IW), strip map (SM), extra wide swath (EW), and wave (WV). As side-viewing devices, all SAR sensors illuminate the imaged ground surface from various view and incidence angles. Radiometric variations caused by local incidence angle may still exist after preprocessing, regardless of preprocessing to correct the effects of incidence angle and topography [5]. IW imagery is collected at incidence angles ranging from 31° to 46°. Sweeps covering these angles are consistently used to collect IW pictures. Sentinel 1 delivers 6-day repeat imagery over Europe. Operational IW images are acquired at nominal 6-day intervals over Europe and 12-day intervals over the rest of the Earth's surface. Higher repeat frequencies are used at higher latitudes and regions where focused acquisitions are intended.

SAR preprocessing is still a technically tricky and computationally demanding undertaking, despite the continuing availability of the Sentinel-1 data and the free, open-source preprocessing software (SNAP toolbox) provided by ESA. The creation of global-scale data products utilizing satellite image time series has been made possible by the Google Earth Engine (GEE), a cutting-edge computing platform released by Google, Inc. [6]. The GEE has been used to conduct global and regional scale investigations of the land surface dynamics, including the cover of forests, surface water [7], populated areas, cropland and soils and other applications. The proposed work aims to develop STanderdized Analysis Ready (STAR) Data in GEE. Sentinel-1 level 1 images have already gone through the orbit file application, radiometric correction, and thermal noise removal to produce products with a backscattering coefficient (σ°) in decibels (dB). STAR includes various corrections -

Border Noise, Lee sigma, terrain, slope, and incident angle- to produce standardized Sentinel 1 level 1 Ground Range Detected (GRD) data. As mentioned earlier, the processing is performed on both the VV and VH bands of the level 1 GRD product. GRD images are fetched using JavaScript code specifying the data collection, satellite, and instrument parameters (like Orbit, Polarization and swath), and the date and the area of interest are used to filter the data further [8].STAR can be used in any application like mapping, such as Forestry [9], Agriculture [10], and Disaster Management (Flood, Oil Spill, and other applications) [11]. Researchers and decision-makers can use the STAR data cube to prepare on-fly flood maps through GEE API services. Such analysis-ready data cube allows for preparing flood maps with better accuracy. The data cube will help keep track of flood occurrence and intensity throughout the years, make near real-time damage assessments, and plan intervention and relief work. Thus, the proposed cloud-based data cube will be helpful for anticipatory and early action during a flood event to provide early support to the affected community.

## II. METHODOLOGY

The Sentinel 1 SAR data products consist of 3 levels: Level 0, Level 1and Level 2. Level 0 SAR products are the raw data directly downlinked from the satellite [12]. In GEE, only Level 1 GRD products are available. The Level 1 GRD product in GEE is prepared by first applying orbit files containing precise satellite tracks, Thermal noise removal and Radiometric calibration with Terrain correction using SRTM 30 or ASTER DEM (use of other DEMs is possible, and user can also use DEMs generated by themselves for this operation) for areas greater than 60 degrees latitude [13]. In order to account for the change in the Earth's curvature, it is essential to adjust the sample start time while creating level-1 products. Azimuth and range compression also produce radiometric artifact at the edges of the images. Additive thermal noise, especially in the cross-polarization channel, degrades Sentinel-1 image intensity [14]. In particular, thermal noise removal minimizes noise effects in the inter-sub-swath texture, normalizing the backscatter signal over the whole Sentinel-1 image and lowering discontinuities between sub-swaths for scenes in multi-swath acquisition modes. Calibration is the procedure that converts digital pixel values to radiometrically calibrated SAR backscatter. The information required to apply the calibration equation is included within the Sentinel-1 GRD product. The DN values of the data products need to be converted to sigma nought values for processing and applying various preprocessing steps further down the line. sigma represents the radar cross-section of a distributed target over that expected from an area of one square meter. It determines the strength of reflection in terms of the geometric cross-section of a conducting sphere. The sigma nought significantly varies with the incidence angle, wavelength, polarization, and scattering surface characteristics. The terrain-corrected values are converted to decibels via log scaling ($10*\log10(x)$).

A standard workflow is shown to build a STAR data cube in the GEE platform (**Figure 1**). The GEE processing workflow comprises a few steps designed to reduce as much error propagation as possible to the subsequent processes down the line.

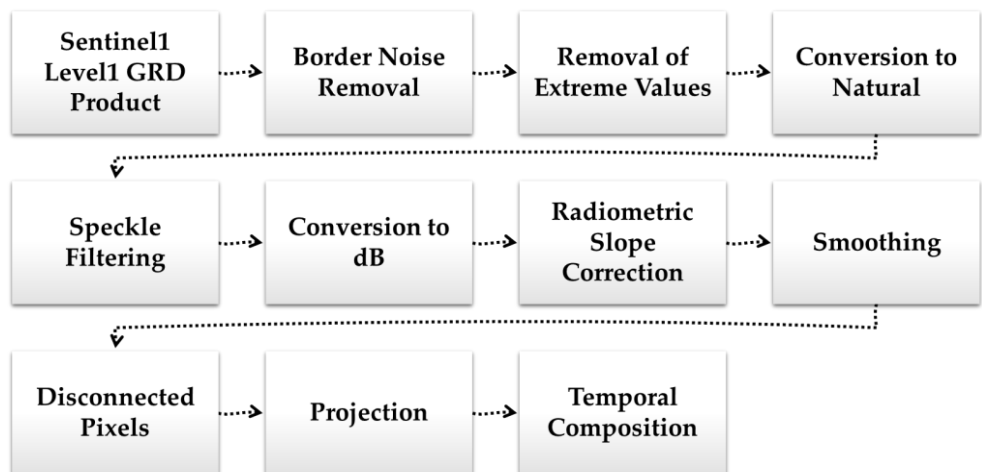

**Figure 1** *Sentinel-1 Ground Range Detected (GRD) processing workflow in GEE*

*1)* **Border Noise Removal**

The acquisition of the geometry of the ascending and descending trajectory over a similar area causes the border noise, which is an artifact. Processing issues may be brought up by the border between acquisitions [15]. Sentinel-1 GRD products contain border noise, an unintended processing artifact that prevents their full use in many applications. This specific form of noise affects over 800,000 GRD products produced before March 2018 [16]. The border noise can cause misclassification of the pixels at the edges of an image. The noise removal is done using the GEE platform to write a user-defined function, which takes the "Angle" values from the band and masks the extreme edges of the image where the angle values are outside the range defined by the user.

*2)* **Removal of extreme values**

Extreme backscatter values are noise caused by the multiple scatterings of the signal with surface structures. This causes extreme backscatter values, which can be seen as white spots or missing pixels in the image [17]. Extreme sigma nought values are excluded by providing the function with threshold values and making a masking function that can be mapped over an Image collection in GEE. The resulting masking layer filters those extreme values from the data. The masking uses the available *updateMask()* function in the GEE code editor.

*3)* **Speckle Filtering**

Wave interference from numerous elementary scatterers causes speckles, which appears as granular noise in SAR images [18]. Reducing speckles improves image quality through the process of speckle filtering. Speckle is not propagated in subsequent operations when such a method is carried out at an early stage of processing SAR data. Since it could eliminate nuanced information about small spatial structures, speckle filtering is not advised when trying to discern small spatial structures or image texture. There are various types of filtering widely used. These are Lee, Improved Lee, Lee Sigma, Gamma, Multi and Mono temporal speckle filters.

Lee filter is a standard deviation-based filtering technique. The process is done using moving windows over the input image. The input to the Lee sigma speckle filtering operation must be in the natural form, *i.e.,* the pixel value must be integer values. The function cannot be applied to the float values of the product's sigma nought values. The GRD products in GEE platforms are in sigma nought (dB) values. This requires converting the data to natural and, after the processing, converting it back to dB values. This is also done by writing a function in GEE that log scales the values. The function is mapped over the image collection.

The Refined Lee filter is a modification of the Lee filter. More recently, multi-temporal speckle filters have been developed to reduce speckle, taking advantage of multiple SAR observations [19]. The function is written for applying the filter in GEE, which uses kernels of defined sizes (3X3, 7X7) and the in-build function of the neighbourhood reducer. The function is then mapped on the Image collection to get the speckle-filtered data product. Enhanced/refined Lee filters reduce speckle noise in radar images while concurrently maintaining texture data. Local statistics (coefficient of variance) is used within each filter window. Each pixel in this is split into three classes, which are then treated as follows- (i) Homogeneous (the average of the filter window repeats the same pixel value), (ii) Heterogeneous (a weighted approximation repeats the pixel value), and (iii) Point target (No modification is made to the pixel value) [20, 21]. Due to many problems in the Lee sigma filter, the new Improved Lee sigma filter is developed. In the Lee sigma filter, pixels within the sigma range are selected from a moving window or kernel, the user defines the size, and the mean value of the selected pixels replaces the centre pixel. However, the image could be over-filtered for a large sigma value and a large moving window size (*i.e.,* $\xi = 0.9$ and a $9 \times 9$ window), causing loss of spatial resolution, which is undesirable for some applications [22].

The *GAMMA filter* is also known as the Maximum A posterior (MAP) filter. This is similar to the Kuan Filter but assumes a gamma distribution [23]. Mono-temporal speckle filtering is applied on each of the images in an Image collection separately. The speckle filters can range from the simplest moving window-based boxcar filter to the much more advanced Lee filter [24], Gamma MAP filter [25], Improved Lee sigma filter [26], and Refined Lee filter [26, 27]. Multi-temporal speckle filtering can be made more dependable by adjusting parameters like the number of images utilized in the framework. The framework must be employed regardless of the acquisition orbit to prevent quality reduction of the collection of speckle-filtered Sentinel-1 images [28].

*4)* **Radiometric slope correction**

Images with some distortion associated with side-looking geometry are produced by SAR data, which is typically detected with a variable viewing angle higher than 0 degrees. In order to make the geometric representation of the image as accurate to reality as possible, terrain adjustments are designed to account for these aberrations. Range Doppler terrain correction uses a digital elevation model to adjust the location of each pixel to correct geometric distortions brought on by topography, such as foreshortening and shadows. For Radiometric Slope Correction, the satellite DEM (Digital Elevation Model) and a custom function are applied over the Image collection to get the terrain-corrected product. The custom function written in GEE takes the terrain information from the DEM and uses it to "flatten" the terrain in the radar image. This eliminates the shadows in the radar image. This is very useful in applications related to water delineation, specifically in the case of flood inundation mapping. Due to terrain effects, dark, shadowy pixels can be erroneously classified as water pixels. Terrain correction eliminates that possibility.

Slope correction is necessary when working with SAR data-derived water bodies or flood maps in high-elevation regions. The slope correction minimizes the false positives due to residual errors introduced by the terrain correction stage. The errors are caused by inaccuracies in DEM data used in the terrain correction process. Steep slope masking can be applied to the generated layers using a DSM (Digital Surface Model) to eliminate these errors. The DSM is used to get the gradient information of the scene, and the user-specified angle value is used to mask the values [29].

*5)* **Smoothing**

Smoothing is a kernel operation that is used to smooth the variations in an image. Smoothing acts as a lowpass filter, filtering out sharp tonal changes in an image. Low pass filtering involves a specific size and shape of the kernel, and the average or median of pixels are calculated, and the newly transformed values are used to get the smoothing effect. The smoothing methods include the *convolve()* function in the GEE platform. This can be used in a function, and the user can define the kernel's size and shape (Square, Circle) to vary the resulting amount of smoothing [30]. The smoothing function smoothes and brings out the separation between the sharp tonal values of an image, *i.e.,* the separation of dark water bodies and the lighter land surface.

*6)* **Disconnected pixels**

Calculate the number of pixels composing objects using the *connectedPixelCount()* image method. Knowing the number of pixels in an object can be helpful for masking objects by size and calculating object area. The following snippet applies *connectedPixelCount()* to the "labels" band of the objectId image defined in the previous section.

*7)* **Projection**

The GRD image collection contains data from two satellite platforms in the Sentinel-1 constellation (Sentinel-1a and Sentinel-1b). These satellites orbit around 180° apart from each other and have a combined global revisit time of 6 days. There is scene overlap for locations away from the equator, and areas may be observed more often than every 6 days but from different orbit paths. The images from different paths will not align with each other. Furthermore, even images from the same relative path (taken 6 or 12 days apart) will not align with each other. The following code demonstrates this by comparing two scenes taken 12 days apart. The scenes share the same relative orbit; however, the orbits and the alignment of observations are not exactly the same.

In practice, some areas belong to different orbits. The variations of incidence, of the order of degrees, can generate variations of projections. During terrain correction, imagery is needed to project to WGS84 latitude/longitude projection (e.g., EPSG:4326). In GEE, projection consists of a CRS and a Scale.

*8)* **Temporal Composition**

Equal-interval synthesis can be used to obtain seamless composite images of a large-scale regional time series, and it can provide data support for classification tasks and other work. Flood mapping of large areas using sentinel-1 in GEE is challenging due to the lack of spatial continuity due to the acquisition strategy. Temporal aggregation—the use of metrics (*i.e.,* mean or median) derived from satellite data over a period of time—is an approach that provides an opportunity to produce a seamless flood mapping. This enables the efficient use of sentinel-1 data. Temporal aggregation is a promising tool for efficiently integrating large amounts of data and can compensate for the spatial gap-filling of the targeted area.

The other important factor is the bands of the SAR data being used for the application, *i.e.,* whether the VV band or the VH band is being used can vary results, and the bands are appropriate for different use cases. For applications in flood mapping, VV gives better results as VV is sensitive to surface roughness while VH is sensitive to the canopy of vegetation. VH captures the volumetric reflections better. Volumetric scattering is best when characterizing forest structures, crop type, and their growth cycle. Various band combinations (VV+VH, VV-VH, RVI, etc.) can also be constructed to increase the inter-class separability of different land cover classes [24].

The correction steps (**Figure 2**) described in this paper can be applied to various applications. For example, the smoothed final output of the STAR data cannot be used in Forestry or agriculture because the smoothing reduces the minor variations in the data, *i.e.,* it reduces the granularity of the data, and minor variations caused by the differences in the land cover cannot be detected [16, 31]. Meanwhile, the smoothed output can be used to broadly classify the land and water bodies. The smoothing brings out the inter-class variations of the water and the land surface [21]. For forest applications, radiometric terrain normalization is a suggested preprocessing step. Radiometric terrain normalization is typically not carried out [32] due to the lack of trustworthy and current digital surface models. Terrain correction (orthorectification) can be done using SRTM30 or ASTER DEM for regions located at more than 60◦ latitude [29].

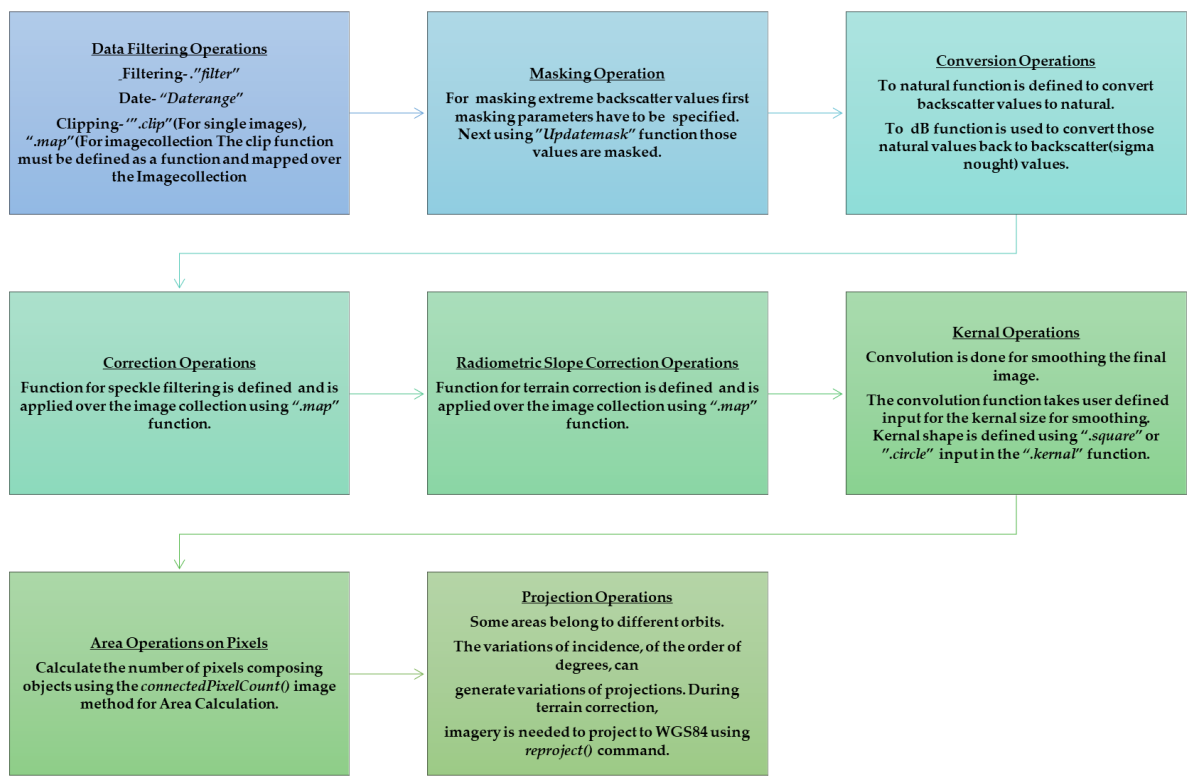

**Figure 2** *Functional flowchart of STAR*

## III. FLOOD MAPPING

The STAR data cube can be used to create a precise flood map. The Otsu algorithm generates an inundation map by extracting a bimodal histogram from pictures [39]. This algorithm improves the interclass separability between the classes of water and non-water. For this investigation, we employed a spatial resolution of 10 m. The algorithm then divides the image into a grid, using chessboard segmentation, and checks each subregion for a bimodal histogram using the highest normalized between-class variance. It must be calculated, and an initial estimate of the segment's water/non-water content must be given in order to estimate the probabilities of the various classes [29]. The STAR data cube has been used to estimate flood extent area in a stretch of Benue River, Nigeria during 2022 flood. For the mentioned stretch of the river, the water extent increased 120.54 sq km from July 2022 to 21 September 2022, with a significant flood of 154.04 sq km on 21 September 2022, including permanent water bodies (**Figures** 3a, 3b and 3c).

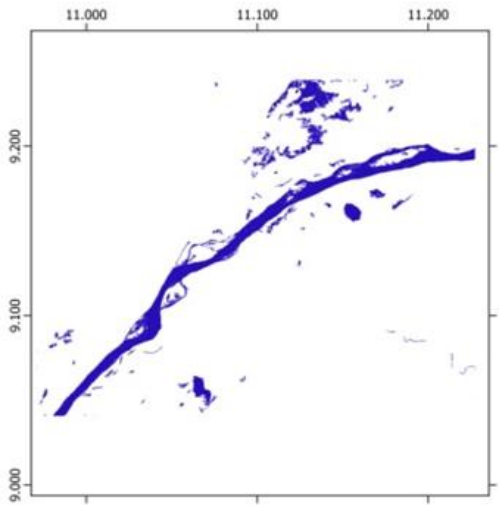
**Figure 3a** *Pre-flood water extent (July 2022)*

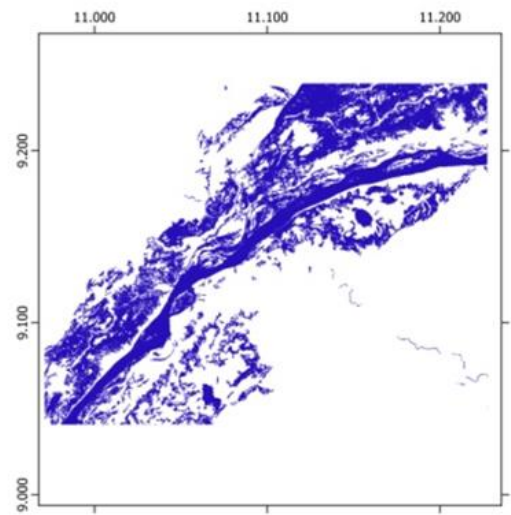
**Figure 3b** *During flood water extent (21/09/2022)*

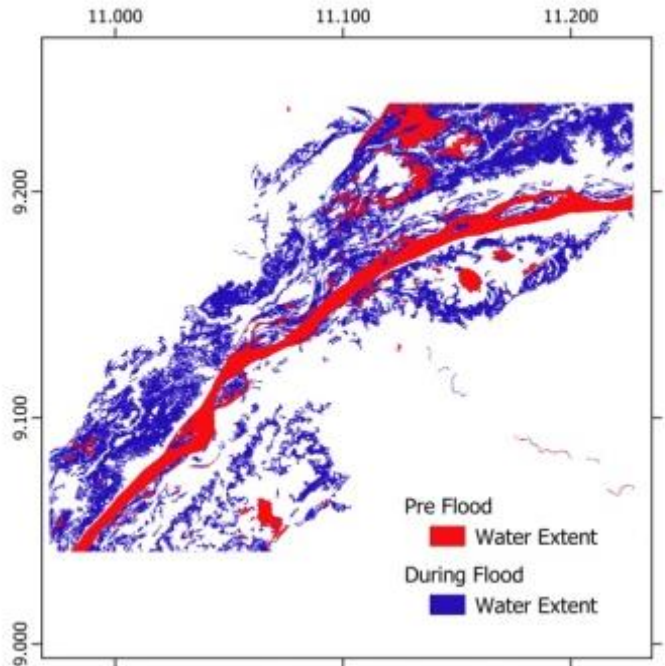
**Figure 3c** *Flood extent (21/09/2022)*

## IV. ACKNOWLEDGEMENTS

This work has been prepared as an output of Digital Innovation for Water Secure Africa (DIWASA) supported by The Leona M. and Harry B. Helmsley Charitable Trust.

## V. REFERENCES

[1] A. Khouakhi, J. Zawadzka, and I. Truckell, "The need for training and benchmark datasets for convolutional neural networks in flood applications", *Hydrology Research*, IWA Publishing, pp 795–806, 2022.

[2] K.N. Markert, A.M. Markert, T. Mayer, C. Nauman, A. Haag, A. Poortinga, B. Bhandari, N.S. Thwal, T. Kunlamai, F. Chishtie, and M. Kwant, "Comparing sentinel-1 surface water mapping algorithms and radiometric terrain correction processing in southeast asia utilizing google earth engine", *Remote Sensing*, MDPI, 12(15), p.2469, 2020.

[3] S. Ghosh, and J. Mukherjee, "Earth observation data to strengthen flood resilience: a recent experience from the Irrawaddy River", *Natural Hazards*, Springer, pp.1-6, 2022.

[4] Sarkar, S., "Drought and flood dynamics of Godavari basin, India: A geospatial perspective", *Arabian Journal of Geosciences*, Springer, 15(8), pp.1-15, 2022

[5] Small, D., 2011. Flattening gamma: Radiometric terrain correction for SAR imagery. *IEEE Transactions on Geoscience and Remote Sensing*, *49*(8), pp.3081-3093.

[6] Pulvirenti, L., Pierdicca, N., Chini, M. and Guerriero, L., "An algorithm for operational flood mapping from Synthetic Aperture Radar (SAR) data using fuzzy logic", *Natural Hazards and Earth System Sciences*, Copernicus Publications, 11(2), pp.529-540, 2011.

[7] Donchyts, G., Baart, F., Winsemius, H., Gorelick, N., Kwadijk, J. and Van De Giesen, N., 2016. Earth's surface water change over the past 30 years. *Nature Climate Change*, *6*(9), pp.810-813.

[9] Nicolau, A.P., Flores-Anderson, A., Griffin, R., Herndon, K. and Meyer, F.J., 2021. Assessing SAR C-band data to effectively distinguish modified land uses in a heavily disturbed Amazon forest. International Journal of Applied Earth Observation and Geoinformation, 94, p.102214.

[10] Liu, C.A., Chen, Z.X., Yun, S.H.A.O., Chen, J.S., Hasi, T. and PAN, H.Z., 2019. Research advances of SAR remote sensing for agriculture applications: A review. Journal of integrative agriculture, 18(3), pp.506-525.

[11] Singha, S., Bellerby, T.J. and Trieschmann, O., 2012, July. Detection and classification of oil spill and look-alike spots from SAR imagery using an artificial neural network. In 2012 IEEE International Geoscience and Remote Sensing Symposium (pp. 5630-5633). IEEE.

[12] https://sentinels.copernicus.eu/web/sentinel/technical-guides/sentinel-1-sar/products-algorithms/level-0-products (Last accessed on 19th February 2023)

[13] https://sentinels.copernicus.eu/web/sentinel/technical-guides/sentinel-1-sar/products-algorithms/level-1-algorithms/ground-range-detected (Last accessed on 19th February 2023)

[14] https://sentinels.copernicus.eu/web/sentinel/user-guides/sentinel-1-sar/overview (Last accessed on 19th February 2023)

[15] Luo, Y. and Flett, D., 2018, March. Sentinel-1 Data Border Noise Removal and Seamless Synthetic Aperture Radar Mosaic Generation. In *Proceedings* (Vol. 2, No. 7, p. 330). MDPI.

[16] Stasolla, M. and Neyt, X., 2018. An operational tool for the automatic detection and removal of border noise in Sentinel-1 GRD products. *Sensors*, *18*(10), p.3454.

[17] Rana, V.K. and Suryanarayana, T.M.V., 2019. Evaluation of SAR speckle filter technique for inundation mapping. *Remote Sensing Applications: Society and Environment*, *16*, p.100271.[44] Islam, M.T. and Meng, Q., 2022. An exploratory study of Sentinel-1 SAR for rapid urban flood mapping on Google Earth Engine. *International Journal of Applied Earth Observation and Geoinformation*, *113*, p.103002.

[18] Markert, K.N., Markert, A.M., Mayer, T., Nauman, C., Haag, A., Poortinga, A., Bhandari, B., Thwal, N.S., Kunlamai, T., Chishtie, F. and Kwant, M., 2020. Comparing sentinel-1 surface water mapping algorithms and radiometric terrain correction processing in southeast asia utilizing google earth engine. *Remote Sensing*, *12*(15), p.2469.

[19] Le Minh, H., Van, T.V. and Asnh, T.T., 2019, October. Using dual-polarization Sentinel-1A for mapping vegetation types in Daklak, Vietnam. In *Proceedings of the 40th Asian Conference on Remote Sensing (ACRS 2019), Daejeon, Korea* (pp. 14-18).

[20] G. Vasumathi "A Survey on SAR Image Classification" International Journal of Advanced Engineering and Global Technology, Vol. 3, Issue12, Page 1461-1465, 2015,.f.,g

[21] Rana, V.K. and Suryanarayana, T.M.V., 2019. Evaluation of SAR speckle filter technique for inundation mapping. *Remote Sensing Applications: Society and Environment*, *16*, p.100271.

[22] A.V. Meenakshi and V. Punitham "Performance of Speckle Noise Reduction Filters on Active Radar and SAR Images" International Journal of Technology And Engineering System (IJTES), Vol2.No1, Page 111- 114,2011M,mvbm,

[23] A.Rajamani and V.Krishnaveni "Performance Analysis Survey of Various SAR Image Despeckling Techniques" International Journal of Computer Applications, Volume 90 – No 7, Page 5-17, 2014

[24] Lee, J.S. Digital image enhancement and noise filtering by use of local statistics. IEEE Trans. Pattern Anal. Mach. Intell. 1980, 2, 165–168.

[25] Lopes, A.; Nezry, E.; Touzi, R.; Laur, H. Maximum a posteriori speckle filtering and first order texture models in SAR images. In Proceedings of the 10th Annual International Symposium on Geoscience and Remote Sensing, College Park, MD, USA, 20–24 May 1990; pp. 2409–2412.

[26] Lopes, A.; Touzi, R.; Nezry, E. Adaptive speckle filters and scene heterogeneity. IEEE Trans. Geosci. Remote Sens. 1990, 28, 992–1000.

[31] Lee, J.S.; Wen, J.H.; Ainsworth, T.L.; Chen, K.S.; Chen, A.J. Improved sigma filter for speckle filtering of SAR imagery. IEEE Trans. Geosci. Remote Sens. 2008, 47, 202–213.

[27] Lemoine, G. Refined Lee GEE Implementation. Available online: https://code.earthengine.google.com/5d1ed0a0f0417f098fdfd2    fa137c3d0c (Last accessed 01/04/2021).

[28] Mullissa, A., Vollrath, A., Odongo-Braun, C., Slagter, B., Balling, J., Gou, Y., Gorelick, N. and Reiche, J., 2021. Sentinel-1 sar backscatter analysis ready data preparation in google earth engine. *Remote Sensing*, *13*(10), p.1954.

[29] Marzi, D. and Gamba, P., 2021. Inland water body mapping using multitemporal sentinel-1 sar data. *IEEE Journal of Selected Topics in Applied Earth Observations and Remote Sensing*, *14*, pp.11789-11799.

[30] https://developers.google.com/earth-engine/guides/image_convolutions (Last accessed 16/02/2023)

[31] Filipponi, F., 2019. Sentinel-1 GRD preprocessing workflow. In *International Electronic Conference on Remote Sensing* (p. 11). MDPI.

[32] Mullissa, A., Vollrath, A., Odongo-Braun, C., Slagter, B., Balling, J., Gou, Y., Gorelick, N. and Reiche, J., 2021. Sentinel-1 sar backscatter analysis ready data preparation in google earth engine. *Remote Sensing*, *13*(10), p.1954.